# Three-dimensional single gyroid photonic crystals with a mid-infrared bandgap

[1]Siying Peng, [2]Runyu Zhang, [1]Valerian H. Chen, [1]Emil T. Khabiboulline, [2]Paul Braun, [1]Harry A. Atwater

1. Applied Physics, California Institute of Technology

2. Department of Materials Science and Engineering, University of Illinois at Urbana-Champaign

**ABSTRACT:** A gyroid structure is a distinct morphology that is triply periodic and consists of minimal isosurfaces containing no straight lines. We have designed and synthesized amorphous silicon (a-Si) mid-infrared gyroid photonic crystals that exhibit a complete bandgap in infrared spectroscopy measurements. Photonic crystals were synthesized by deposition of a-Si/$Al_2O_3$ coatings onto a sacrificial polymer scaffold defined by two-photon lithography. We observed a 100% reflectance at 7.5 μm for single gyroids with a unit cell size of 4.5 μm, in agreement with the photonic bandgap position predicted from full-wave electromagnetic simulations, whereas the observed reflection peak shifted to 8 μm for a 5.5 μm unit cell size. This approach represents a simulation-fabrication-characterization platform to realize three-dimensional gyroid photonic crystals with well-defined dimensions in real space and tailored properties in momentum space.

**TABLE OF CONTENTS GRAPHIC**

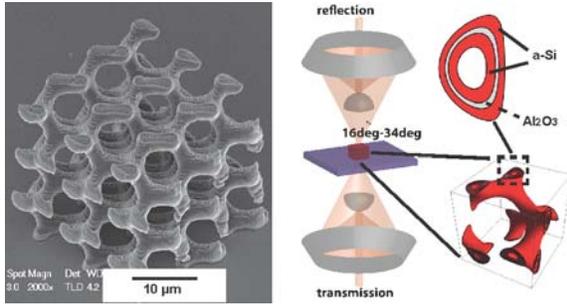

Three-dimensional photonic crystals offer opportunities to probe interesting photonic states such as bandgaps[1-8], Weyl points[9, 10], well-controlled dislocations and defects[11-13]. Combinations of morphologies and dielectric constants of materials can be used to achieve desired photonic states. Gyroid crystals have interesting three-dimensional morphologies defined as triply periodic body centered cubic crystals with minimal surfaces containing no straight lines[9, 10, 14-19]. A single gyroid structure, such as the one shown in Fig. 1a, consists of isosurfaces described by

$$Sin(x)Cos(y) + Sin(y)Cos(z) + Sin(z)Cos(x) > u(x,y,z),$$

where the surface is constrained by $u(x, y, z)$. Gyroid structures exist in biological systems in nature. For example, self-organizing process of biological membranes forms gyroid photonic crystals that exhibit the iridescent colors of butterfly's wings[20]. Optical properties of gyroids could vary with tuning of $u(x,y,z)$[21], unit cell size, spatial symmetry[9] as well as refractive index contrast. Single gyroid photonic crystals, when designed with high refractive index and fill fraction, are predicted to possess among the widest complete three-dimensional bandgaps[9, 22], making them interesting for potential device applications such as broadband filters and optical cavities. In this work, we demonstrate a synthesis approach for forming mid-infrared three-dimensional gyroid photonic crystals, and report experimental measurements of the bandgap for a single gyroid structure at mid-infrared wavelengths.

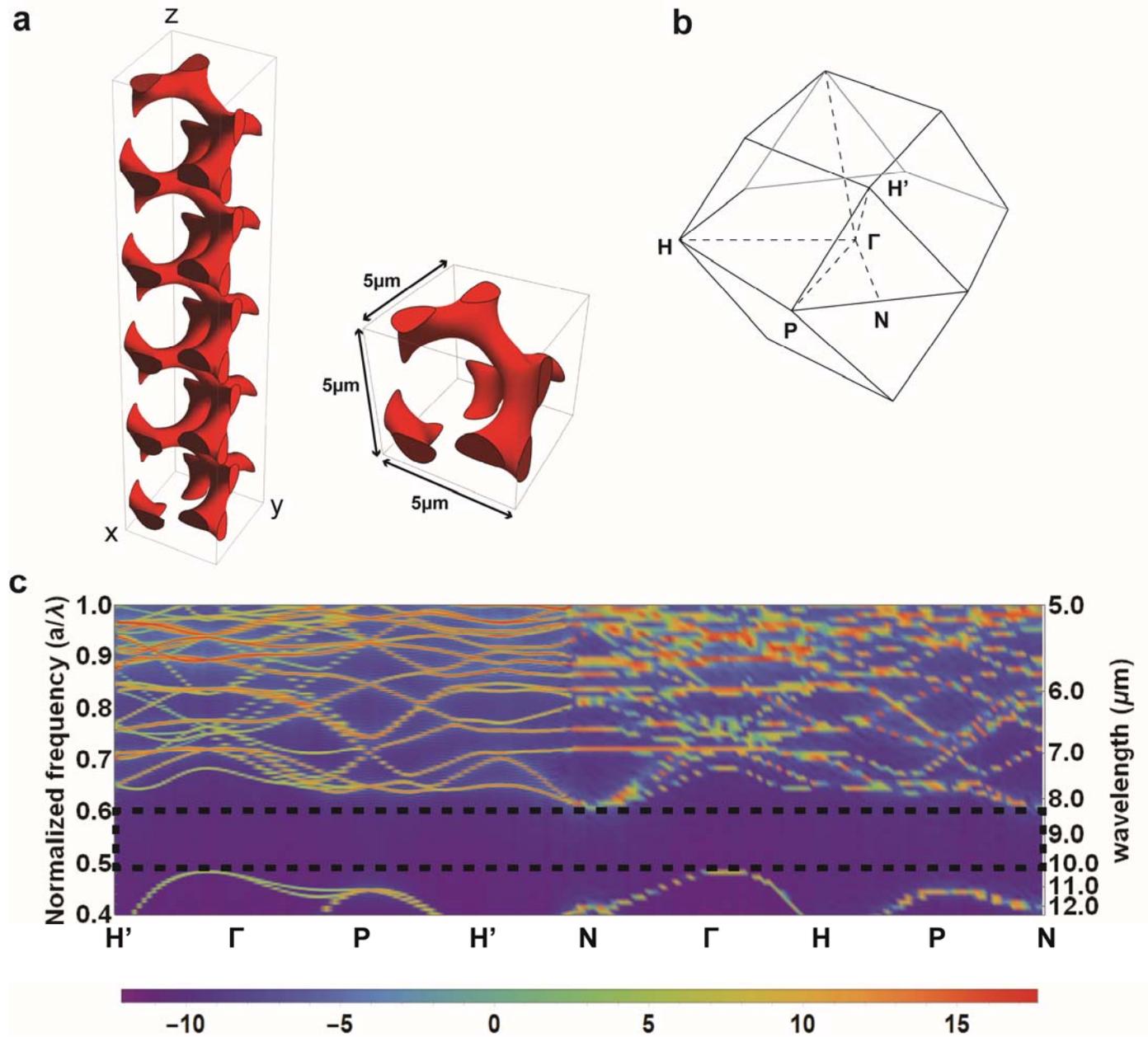

**Figure 1 Single gyroid structure and photonic bandstructure simulation.** (a)Stacks of unit cells of single gyroid structure (b) bcc Brillouin zone. (c) photonic band structure of a-Si single

gyroid from full wave simulations, with unit cell size of 5 μm in x, y, z directions. The color map is proportional to logarithm of the density of states.

**RESULTS**

To realize gyroid photonic crystals at mid-infrared wavelengths, we utilize full-wave finite-difference time-domain (FDTD) simulations to determine the dimensions and materials required for crystal design (see Methods). The simulation shown in Fig. 1c reveals that a-Si single gyroid crystals with a unit cell size of 5μm and u(x,y,z)=1.1 (see Table S1 for fill fraction) has a complete bandgap (indicated by the dashed box) from 8μm to 10μm in all symmetry directions of the bcc Brillouin zone in Fig. 1b. In units of normalized frequency calculated by dividing unit cell size (a) by wavelength (λ), the complete bandgap is between 0.5 and 0.6. For constant refractive index at mid-infrared wavelength, we can use the value of normalized frequency to deduce bandgap position for crystals with different unit cell size. For example, for a=5.5μm single gyroid crystal investigated in Fig.4b, the center of the bandgap can be inferred by a normalized frequency, resulting in a shift of the bandgap center by 0.6 μm. Therefore, we identified a-Si as a suitable material with its high refractive index and low loss at mid-infrared wavelength. Another suitable candidate material is germanium, which has even higher refractive index (n>4) and sufficiently low loss in this wavelength regime.

To fabricate a-Si single gyroid structures, we developed a protocol which incorporates multiple steps[2, 12, 13, 23, 24] (see Methods for detailed description), as illustrated in Fig. 2. Two-photon lithography was utilized to directly write a sacrificial polymer scaffold of gyroid photonic crystals with unit cell sizes of 4.5, 5.1 and 5.5 μm on mid-infrared transparent silicon substrates. Each sample is composed of 20x20x10 unit cells. We conformally deposited 40 nm thick aluminum oxide coatings on the polymer gyroids via atomic layer deposition (ALD) at $150^0$C. We then used focused ion beam (FIB) milling to remove the crystal sides to facilitate polymer removal, yielding a hollow inorganic aluminum oxide crystal after oxygen plasma cleaning. Subsequently, the structure was conformally coated and in-filled with a 100nm a-Si layer at $350^0$C using chemical vapor deposition (CVD). The polymer structure is not structurally stable at temperatures above $250^0$C, which are typically necessary for conformal deposition of high refractive index materials such as a-Si. Therefore the hollow aluminum oxide crystal is a critical intermediate structure to provide a scaffold that can withstand high temperature. The final structure consists of a 40nm middle layer of aluminum oxide and two 100nm/150nm a-Si layers on both the inside and outside of the aluminum oxide scaffold, corresponding to u(x,y,z) values of 1.1/1.05, 1.2, 1.25 and 1.35/1.37 for coated a-Si, $Al_2O_3$, in-filled a-Si and inner hollow part respectively (see Table S1 for fill fraction values).

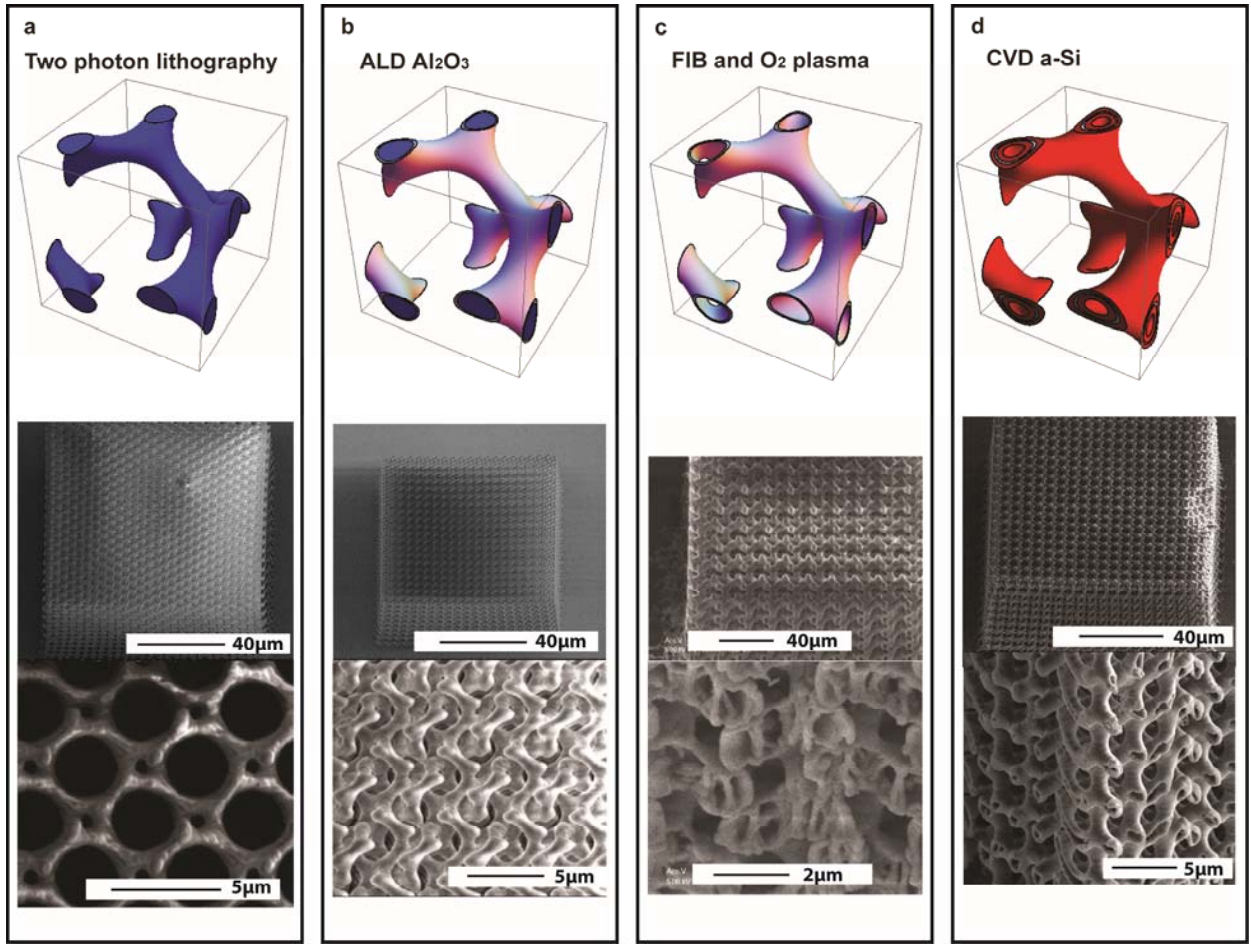

**Figure 2 Fabrication procedures of gyroid photonic crystals.** (a) two photon lithography to define a sacrificial polymer scaffold of the gyroid structure. (b) atomic layer deposition of $Al_2O_3$ to coat the polymer structure. (c) focused ion beam milling to remove the sides of the structure, followed by oxygen plasma to remove the polymer, leaving a hollow alumina structure. (d) chemical vapor deposition of a-Si to coat and in-fill the hollow alumina structure.

We characterized the resulting a-Si single gyroid photonic crystals by Fourier transform infrared spectroscopy (FTIR), shown in Fig. 3a (see Methods for a detailed description of the measurements). SEM images of the characterized sample are shown in Fig. 3b. The reflectance spectrum of the a=4.5μm sample reveals a peak of 98% at 7.0 μm as shown in Fig. 3c (red dashed line), in agreement with the predicted reflectance peak center of the 4.5μm trapezoidal structure in the figure above (black dashed line) (see the section of "Full wave simulation on deformed crystals" in Methods), as well as the band gap center for a cubical single gyroid structure (gray dashed line). The reflectance of the sample is normalized to the reflectance of an atomically smooth gold mirror of 97% reflectance. Additional 50nm coating and in-filling of a-Si on the structure red shifted the reflectance peak to 7.5 μm, giving rise to the 100% reflectance peak shown in Fig. 3c (red solid line). The reflectance peak at 7.5μm is a direct manifestation of a photonic bandgap. The transmittance spectrum shown in Fig. 4a has a wide 0% transmittance band centered at 7.5 μm, confirming the bandgap. The extinction (scattering+absorption) is obtained by subtracting the transmittance and reflectance percentages from 100%. Since the FTIR collection angle is limited to $16^0$-$34^0$, that part of the reflected light that lies outside of this angular range is considered as scattering here. These results reveal the photonic property of the single gyroid structure, namely the optical bandgap in the mid-infrared regime.

We also characterized the reflectance spectrum of the a=5.1 μm period and a=5.5 μm period samples and compared them with that of the a=4.5μm sample, as shown in Fig. 4b. We observed a red shift of the reflection peaks by 0.6μm in wavelength for the 5.5μm period sample, and a shift of 0.4 μm in the 5.1μm sample relative to the 4.5μm sample, which is in agreement with the increase of the bandgap wavelength predicted (see Fig. 1c and Fig. 5). This observation can be intuitively understood by considering the resonant coupling between the incoming beam and the cavity of a periodic unit cell. The agreement between our experimental results and simulations for three structures with different unit cell sizes confirms the mid-infrared bandgap feature of the single gyroid photonic crystals.

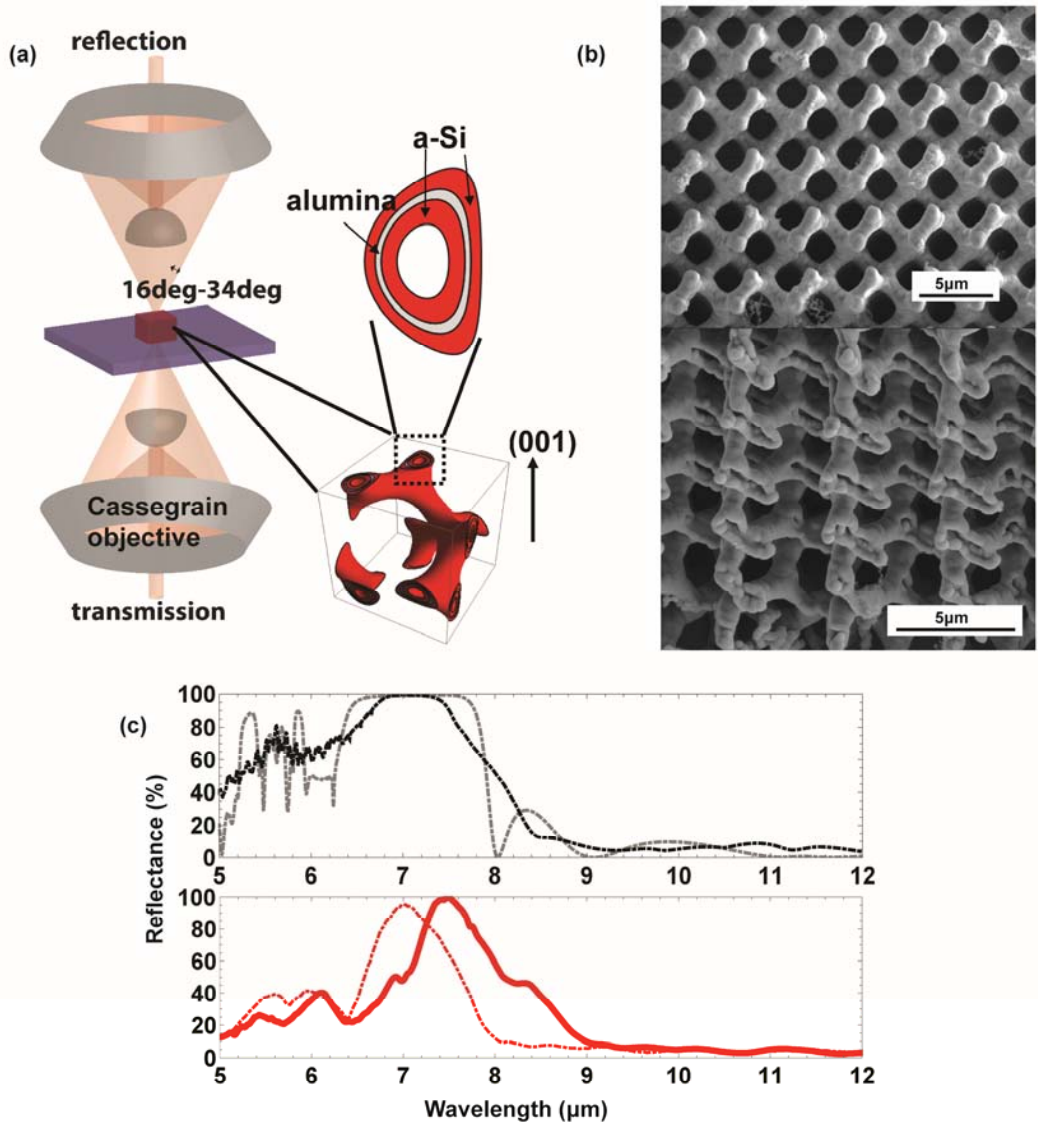

**Figure 3 FTIR characterization of bandgaps.** (a) Fourier transform infrared spectroscopy experimental configuration. (b) SEM images of a hollow gyroid at (001) crystal orientation with a-Si (150nm) /$Al_2O_3$ (40nm) /a-Si (150nm) layers (c) reflectance spectrum from full wave simulations of a hollow gyroid at (001) crystal orientation with a-Si (100nm) /$Al_2O_3$ (40nm) /a-Si

(100nm) layers (gray dashed line), reflectance spectrum from full wave simulations of a trapezoidal hollow gyroid at (001) crystal orientation with a-Si (100nm) /$Al_2O_3$ (40nm) /a-Si (100nm) layers (black dashed line), FTIR measurement of a hollow gyroid with a-Si (100nm) /$Al_2O_3$ (40nm) /a-Si (100nm) layers (red dashed line) and FTIR measurement of a hollow gyroid with a-Si (150nm) /$Al_2O_3$ (40nm) /a-Si (150nm) layers (red dashed line).

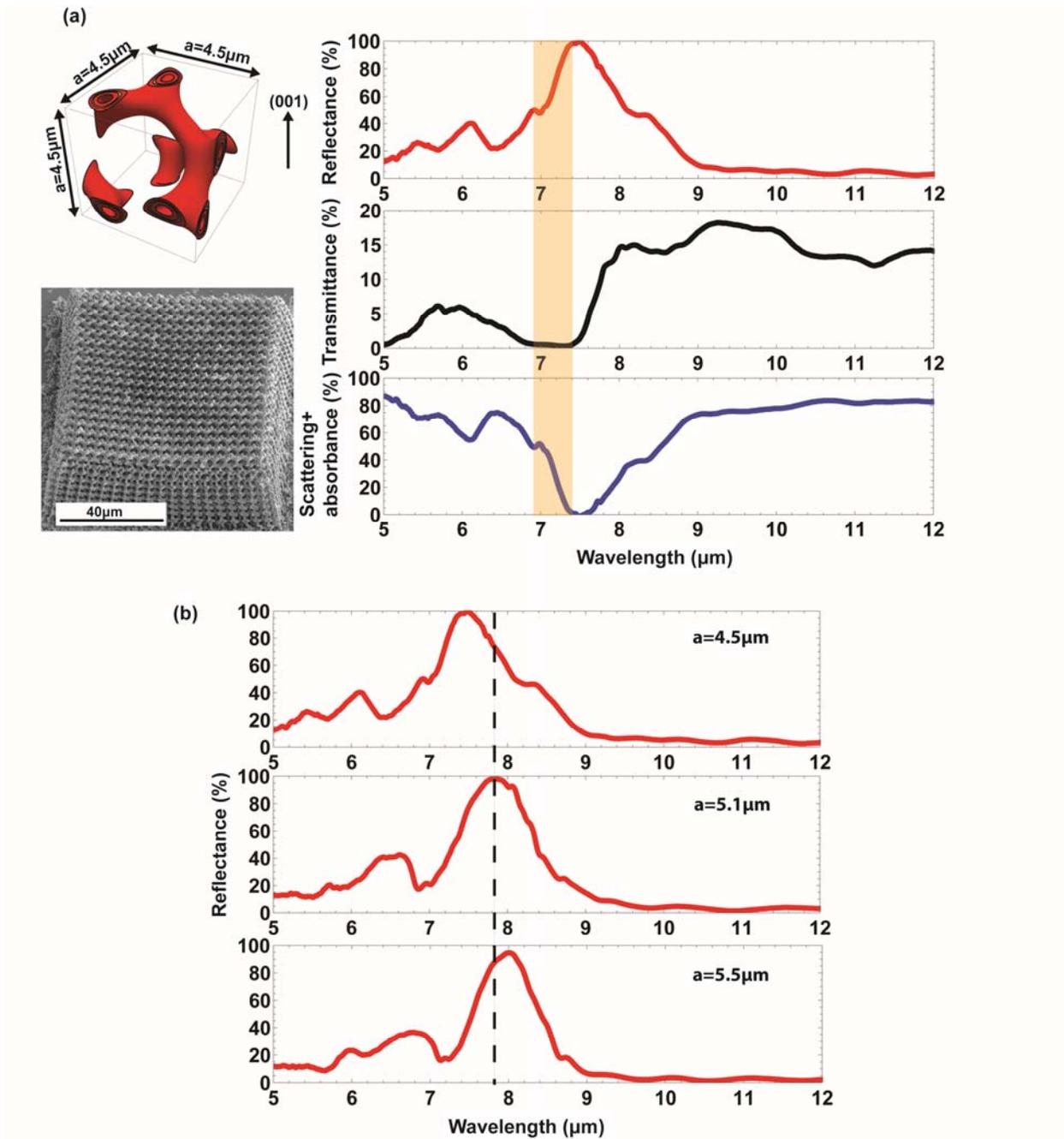

**Figure 4 FTIR characterization of single gyroids.** (a) reflectance, transmittance and scattering+absorbance from a single gyroid structure with unit cell size of 4.5μm and a total of 20x20x10 cells. (b) comparison of reflection spectra of samples with unit cell sizes of 4.5μm, 5.1μm and 5.5μm (20x20x10 unit cells).

**DISCUSSION**

Several specific features arise in the reflectance spectra of these single gyroid samples due practical aspects of the chosen material compositions and defects present in the crystal structures. The bandgap center of the 4.5 μm period structure is shifted from the predicted 8 μm wavelength inferred from Fig. 1c and Fig. 3c (black dashed line) to the experimentally measured 7.0 μm wavelength in Fig. 3c (red dashed line). This is due to lower effective refractive index of the a-Si/$Al_2O_3$/a-Si heterostructure present in fabricated samples as compared with an assumed homogeneous dense solid a-Si cross-section for the photonic crystal elements in the simulations given in Fig. 1c. The center of the bandgap calculated from full wave simulation for the actual a-Si (100nm) /$Al_2O_3$ (40nm) /a-Si (100nm) composition is at 7 μm, shown in Fig. 5. Another minor contribution comes from the lower refractive index in the experimental a-Si as compared to those of Palik[25] used in full wave simulations, shown in Fig. S1a, likely due to differences in the deposition conditions.

The main crystal defect is a overall shape distortion of the lattice which is attributed to polymer shrinkage, as shown in SEM images in Fig. 2 and Fig. 4, instead of the cubic shape expected in an ideal lattice. A trapezoidally shaped crystal is formed as the result of the polymer scaffold shrinkage. After direct laser writing, the written IP-Dip photoresist cross-links to form polymer scaffolds. The unwritten photoresist is then removed in the process of development, during which the top portion of the polymer scaffold shrinks. The adhesion force between the scaffold and the rigid Si substrate prevents of the bottom lattice from shrinking. This distortion results in a decreasing unit cell size from the bottom to the top of the crystal, resulting in an overall trapezoidal shape for the crystal. In an ideal cubic crystal, we expect a 100% reflectance peak with bandwidth matching that of the predicted bandgap, shown in Fig. 3c (black dashed line). A narrowing of the measured reflectance peak in Fig. 3c (red dashed line) is expected when the trapezoidal shape is taken into account in the deformed crystal simulation (see Methods "Full wave simulation on deformed crystals"), together with the a-Si/Al$_2$O$_3$/a-Si material composition. Intuitively, the measured reflectance spectrum is a superposition of the scattering from each crystal layer over the range of momenta accessible by the $16^0$ to $34^0$ range of incident angles of the FTIR configuration shown in Fig. 3a and Fig. 5 (orange lines). As the unit cell dimension increases from top to bottom of the trapezoidal shaped crystal, the reflectance peak center of each layer is increasingly red shifted. The width of the measured reflectance narrows as a result of the effective inhomogeneity and the lowered effective refractive index. The second reflectance peak observed in Fig. 4b, at 6.1 μm for the a=4.5 μm sample, 6.5 μm for the a=5.1 μm sample and 6.8 μm for the a=5.5 μm sample, respectively, can also be explained by this effective inhomogeineity . From reflectance simulations of an ideal a-Si gyroid crystal in Fig. 3c (gray dashed line), distinct photonic bands exist above the band gap, manifested as sharp dips in the reflectance spectrum. The states observed in the reflectance spectra is equivalent of projected band structure on the (001) plane, spanned by Γ-H and Γ-H' symmetry directions. Band structure simulations in Fig. 1c and Fig. 5 indicate photonic bands above the band gap. A trapezoidal shaped crystal gives rise to spectral inhomogeneity, and consequently the photonic band features broadens into one reflectance peak. Despite realistic material compositions and sample defects, the essential physics interpreted from these reflectance spectra is not affected by the above-mentioned nonidealities.

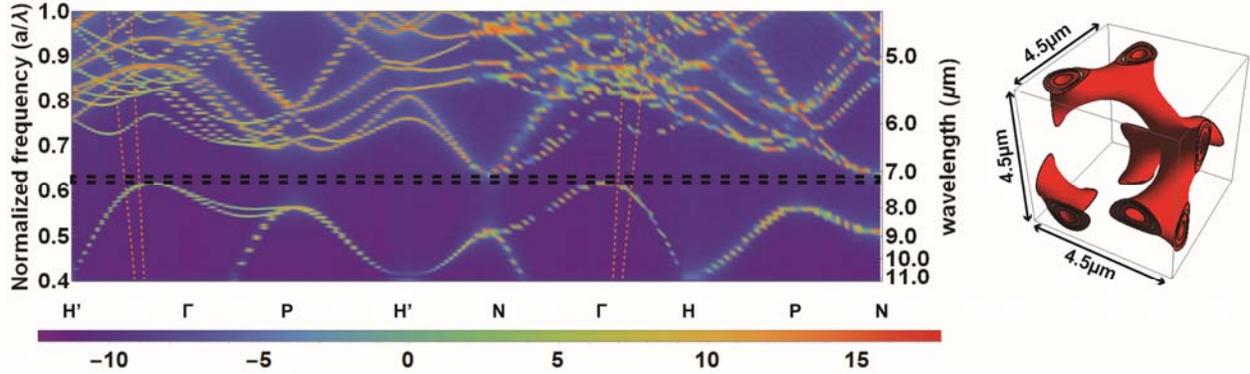

**Figure 5 Photonic band structure simulation.** Photonic band structure of a single gyroid consists of a-Si (100nm) /Al$_2$O$_3$ (40nm) /a-Si (100nm) layers from full wave simulations, with unit cell size of 4.5μm in x, y, z directions. Bands in between the orange lines are accessible through FTIR characterization shown in Fig. 3a

**CONCLUSION**

In conclusion, we experimentally observed a 100% reflective bandgap at mid-infrared wavelengths in single gyroid photonic crystals with high refractive index materials, fabricated using two-photon lithography and conformal layer deposition, confirming photonic bandgap predictions obtained from simulations. This mid-infrared bandgap is also predictably tunable by changing the unit cell size in the simulation design and fabrication. The synthesis/characterization approach described here opens the door to design of more complex mid-infrared photonic crystals with topological states, such as Weyl points in double gyroid photonic crystal with parity-breaking symmetry, for which synthesis of single gyroid photonic crystals establishes feasibility. Further designs may also yield gyroid photonic structures whose surfaces exhibit topologically protected states, suggesting the possibility to synthesize these intriguing structures to create unusual states and phases of light.

**METHODS**

Full wave band structure simulation

Simulations were performed using Lumerical FDTD Solutions v8.15.716. In simulations, bands are excited by randomly placing dipoles inside the simulation region that is defined by Bloch boundary conditions in x, y and z directions. Randomly placed field monitors record electric and magnetic fields over time. Fourier transformation of the overall electric field versus time reveal spectrum of the bands. By tuning phase of the Bloch boundary conditions, we were able to calculate bands at wave vectors along all high symmetry directions in the Brillouin zone of a bcc lattice. Palik[25] n, k data for a-Si are used in the simulation (shown in supplementary Fig. 1S).

Sample fabrication

Polymer gyroid structures were written in negative photoresist IP-Dip using the Photonic Professional GT system (Nanoscribe GmbH). 40 nm thick aluminum oxide coatings on the polymer gyroids were conformally deposited via atomic layer deposition at $150^0C$ in a Cambridge Nanotech S200 ALD System with $H_2O$ and trimethylaluminum (TMA) precursors. We used focused ion beam milling with the FEI Nova 200 Nanolab at 30 kV and 30 nA Ga beam condition to remove the crystal sides to facilitate polymer removal. We etched out the polymers with oxygen plasma using the March PX-500 plasma etcher, yielding a hollow inorganic aluminum oxide crystal. Then the structure is conformally coated and in-filled with 100nm/150nm of a-Si at $350^0C$ using static chemical vapor deposition, with refilled silane as the precursor at an average deposition speed of 10nm/hour. Refractive index of both deposited a-Si and $Al_2O_3$ are measured and shown in supporting information Fig. 1S.

FTIR characterization

The mid-infrared light is incident on the sample at incidence angles from $16^0$ to $34^0$ after being focused with a Cassegrain objective. The sample sits on an intrinsic double side-polished silicon substrate with the (001) crystal surface of the gyroid structure in parallel with the substrate surface. Reflection and transmission spectra are collected from the same range of angles with two identical Cassegrain objectives on each side of the sample respectively. Each incidence angle could excite a corresponding wave vector in a specific symmetry direction in the band structure. For three-dimensional crystal structures, the orientation of the crystal could determine projection of the band structure onto a specific symmetry plane.

Full wave simulation on deformed crystals

Simulations were performed using Lumerical FDTD Solutions v8.15.716. To take into account effects of polymer shrinkage during the structure developing process, we performed full wave simulation approximating the trapezoidal morphology shown in SEM images in Fig. 2 and Fig. 4. The simulated structure is infinitely periodic in the x and y directions and has a finite length of three unit cells in the z direction, with the a-Si/$Al_2O_3$/a-Si material composition. The simulation region has periodic boundary condition in the x and y directions and perfectly matched layer (PML) absorbing boundary condition in the z direction. Plane wave source incidents on the structure in the (001) direction as indicated in Fig. 3a. Frequency domain field monitors are placed above and below the structure to collect reflection and transmission spectra. We repeated the simulation for a series of unit cell sizes ranging from 4.2 μm to 4.7 μm with an increment of 0.1 µm. The arithmetic average of these six reflectance spectra, shown as the black dashed line in Fig. 3c, is used to approximate the trapezoidal effect on the reflectance spectrum.

## ASSOCIATED CONTENT

Supporting Information Available:

Optical constants of deposited materials (a-Si and $Al_2O_3$) and u(x,y,z) versus fill fraction.

## ACKNOWLEDGMENTS

This work is part of the Light Material Interactions in Energy Conversion Energy Frontier Research Center funded by the U.S. Department of Energy, Office of Science, Office of Basic Energy Sciences under Award Number DE-SC0001293.The authors thank George Rossman for FT-IR assistance, the Kavli Nanoscience Institute at Caltech for cleanroom facilities, the Lewis Group ALD facility at Caltech, V. W. Brar and F. Liu for insightful discussions.

**Supporting Information**

**Optical constants of deposited materials**

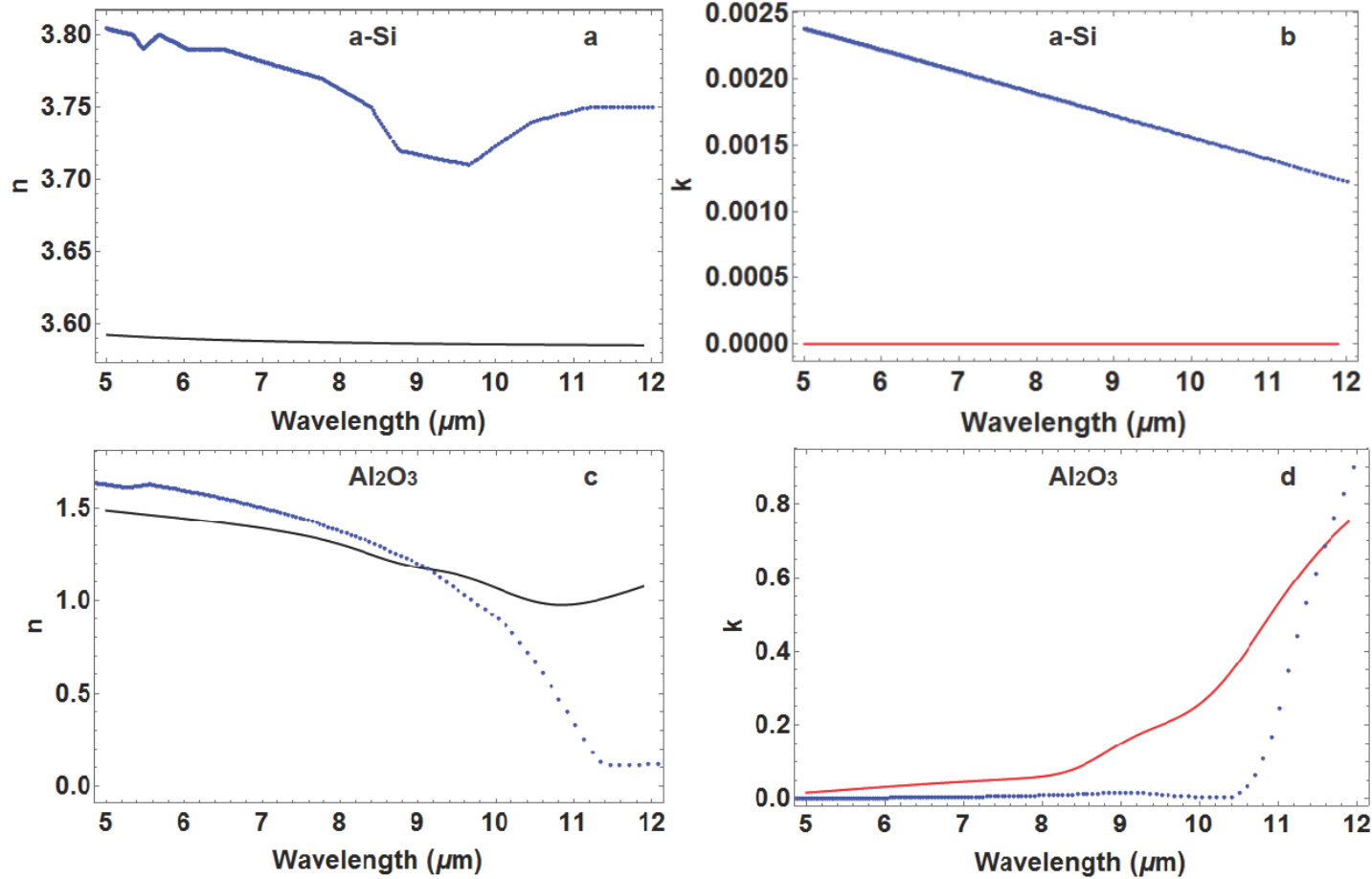

**Figure S1 Optical constants at Mid-infrared wavelength.** (a) measured real part of refractive index of a-Si (black solid line) and Palik's real part of refractive index of a-Si (blue dotted line) (b) measured imaginary part of refractive index of a-Si (red solid line) Palik's imaginary part of refractive index of a-Si (blue dotted line) (c) measured real part of refractive index of $Al_2O_3$ (black solid line) and Palik's real part of refractive index of $Al_2O_3$ (blue dotted line) (d) measured imaginary part of refractive index of $Al_2O_3$ and Palik's real part of refractive index of (blue dotted line)$Al_2O_3$

Psi (Ψ) and delta (Δ) data were measured using IR-VASE Mark II infrared variable angle spectroscopic ellipsometer, from a-Si/$Al_2O_3$ films deposited on an intrinsic silicon substrate. Optical constants n and k were then obtained from psi (Ψ) and delta (Δ) using a three layer fitting model.

**Fill fraction and discontinuity**

| u(x,y,z)      | 1.0   | 1.05  | 1.1   | 1.2   | 1.3   | 1.35  | 1.37  | 1.4   | 1.45  |
|---------------|-------|-------|-------|-------|-------|-------|-------|-------|-------|
| fill fraction | 0.155 | 0.141 | 0.127 | 0.077 | 0.060 | 0.023 | 0.021 | 0.018 | 0.004 |

**Table S1 u(x,y,z) versus fill fraction for a solid single gyroid.**

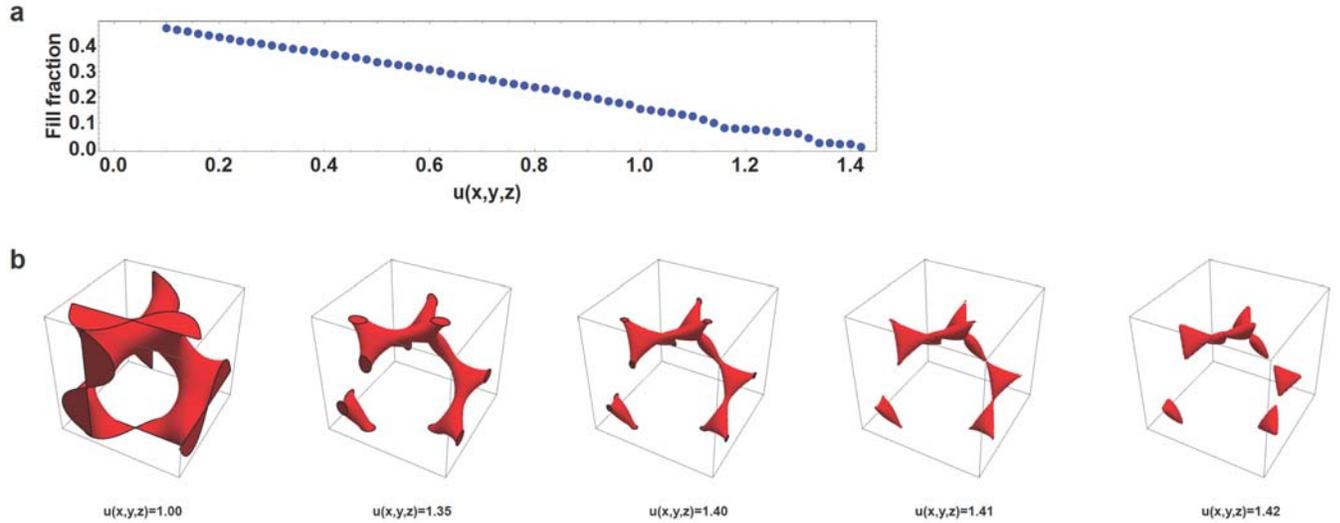

**Figure S2 Single gyroid with different u(x,y,z).** (a) u(x,y,z) versus fill fraction for a solid single gyroid (b) single gyroid with u(x,y,z)=1.00, 1.35, 1.40, 1.41 and 1.42 respectively

Fill fractions are calculated in correspondence to u(x,y,z) for a solid single gyroid structure. Our fabricated structure consists of a-Si (100nm) /Al$_2$O$_3$ (40nm) /a-Si (100nm) is a hollow single gyroid structure with fill fraction of 0.104. The a-Si (100nm) /Al$_2$O$_3$ (40nm) /a-Si (100nm) layers have u(x,y,z) values of 1.1, 1.2 and 1.25 respectively. The inner hollow part corresponds to a connected air gyroid with u(x,y,z)=1.35, shown in Fig. S2b. For hollow single gyroid consists of a-Si (150nm) /Al$_2$O$_3$ (40nm) /a-Si (150nm), the fill fraction is 0.12. The a-Si (150nm) /Al$_2$O$_3$ (40nm) /a-Si (150nm) layers have u(x,y,z) values of 1.05, 1.2 and 1.25 respectively, and an inner air gyroid with u(x,y,z)=1.37. For u(x,y,z)<1.41, a solid gyroid structure is a connected network. The surface becomes disconnected at u(x,y,z)=1.41, as shown in Fig. S2b.